\documentclass[12pt,dvips]{article}
\usepackage{amsmath,amssymb}
\usepackage{jcppre}
\usepackage{epsfig}
\newcommand {\nuc}[2]{\mbox{${}^{#1}\rm #2$}}
\newcommand {\half}{\mbox{$\frac{1}{2}$}}
\newcommand {\smhalf}{\raisebox{0.4ex}{\mbox{$\scriptstyle\frac{1}{2}$}}}
\newcommand{\bra}[1]{\mbox{$\langle{#1}|$}}
\newcommand{\ket}[1]{\mbox{$|{#1}\rangle$}}
\newcommand{\Hz}[1]{\mbox{$\mbox{#1}\,\rm Hz$}}
\newcommand{\MHz}[1]{\mbox{$\mbox{#1}\,\rm MHz$}}
\begin{document}
\title{Implementation of a Quantum Algorithm to Solve Deutsch's Problem
on a Nuclear Magnetic Resonance Quantum Computer}
\author{}
\maketitle
\begin{quotation}
\frenchspacing
\raggedright
\noindent
\mbox{}\\
\begin{singlespace}
J. A. Jones\\
{\sl Oxford Centre for Molecular Sciences, New Chemistry Laboratory,
South Parks Road, Oxford OX1~3QT, UK,} and
{\sl Centre for Quantum Computing, Clarendon Laboratory, Parks Road,
Oxford OX1~3PU, UK}\\
\vspace{0.5in}
M. Mosca\\
{\sl Centre for Quantum Computing, Clarendon Laboratory, Parks Road,
Oxford OX1~3PU, UK,} and
{\sl Mathematical Institute, 24--29 St Giles', Oxford, OX1~3LB, UK}\\
\end{singlespace}
\vspace{0.5in}
Correspondence should be addressed to J.~A. Jones at the New Chemistry
Laboratory.\\
Email: {\tt jones@bioch.ox.ac.uk}\\
\end{quotation}

\begin{abstract}
Quantum computing shows great promise for the solution of many difficult
problems, such as the simulation of quantum systems and the factorisation
of large numbers.  While the
theory of quantum computing is well understood, it has proved difficult
to implement quantum computers in real physical systems.  It has recently
been shown that nuclear magnetic resonance (NMR) can be used to
implement small quantum computers using the
spin states of nuclei in carefully chosen small molecules.  Here we
demonstrate the use of an NMR quantum computer based on the pyrimidine
base cytosine, and the implementation of a quantum algorithm to
solve Deutsch's problem
(distinguishing between constant and balanced functions).  This is the first
successful implementation of a quantum algorithm on any physical system.
\end{abstract}
\newpage

\section{INTRODUCTION}
In 1982 Feynman pointed out that it appears to be impossible to efficiently
simulate the behaviour of a quantum mechanical system with a
computer\cite{feynman1}.
This problem arises because the quantum system is not confined to its
eigenstates, but can exist in any superposition of them, and so the space
needed to describe the system is very large.  To take a simple example,
a system comprising $N$ two level sub-systems, such as $N$ spin-\half\
particles, inhabits a Hilbert space of dimension $2^N$, and evolves
under a series of transformations described by matrices containing
$4^N$ elements.  For this reason it is impractical to simulate the
behaviour of spin systems containing more than about a dozen spins.

The difficulty of simulating quantum systems using classical computers
suggests that quantum systems have an information processing capability
much greater than that of corresponding classical systems.  Thus it
might be possible to build quantum mechanical
computers\cite{feynman1,deutsch1,lloyd,ekert1} which utilise
this information processing capability in an effective way to achieve
a computing power well beyond that of a classical computer.  Such
a quantum computer could be used to efficiently simulate other quantum
mechanical systems\cite{feynman1,lloyd}, or to solve conventional
mathematical problems\cite{ekert1} which suffer from a similar
exponential growth in complexity, such as factoring\cite{shor1}.

Considerable progress in this direction has been made in recent years.
The basic logic elements necessary to carry out quantum computing are well
understood, and quantum algorithms have been developed, both for simple
demonstration problems\cite{deutsch2,deutsch3,ekert2} and for more
substantial problems such as factoring\cite{ekert1,shor1}.
Experimental implementation of a quantum computer has, however, proved
difficult.  Much effort has been directed towards implementing quantum
computers using ions trapped by electric and magnetic fields\cite{cirac},
and while this approach has shown some success\cite{monroe},
it has proved difficult to progress
beyond computers containing a single two-level system (corresponding to
a single quantum bit, or qubit).

Recently two separate approaches have been described\cite{cory2,chuang2}
for the implementation of a quantum computer using nuclear magnetic
resonance\cite{ernst} (NMR).  These approaches shows great promise as it
has proved relatively simple to investigate quantum systems containing two
or three qubits\cite{cory2,chuang2,GHZ}.  Here we describe our
implementation of a simple quantum algorithm for solving Deutsch's
problem,\cite{deutsch2,deutsch3,ekert2} on a
two qubit NMR quantum computer.

\section{QUANTUM COMPUTERS}
All current implementations of quantum computers are built up from a small
number of basic elements.  The first of these is the qubit,
which plays the same role as that of the bit in a classical computer.  A
classical bit can be in one of two states, $0$ or $1$, and similarly a
qubit can be represented by any two level system with eigenstates
labelled \ket{0} and \ket{1}.  One obvious implementation is to use the
two Zeeman levels of a spin-\half\ particle in a magnetic field, and
we shall assume this implementation throughout the rest of this paper.
Unlike a bit, however, a qubit is not
confined to these two eigenstates, but can in general exist in some
superposition of the two states.  It is this ability to exist in
superpositions which makes quantum systems so difficult to simulate and
which gives quantum computers their power.

The second requirement is a set of logic gates, corresponding to gates such
as {\sc and}, {\sc or} and {\sc not} in conventional computers.\cite{feynman2}
Quantum gates differ from their classical counterparts in one very
important way: they must be reversible\cite{feynman2,bennett}.
This is because the evolution of any quantum system
can be described by a series of unitary transformations, which are themselves
reversible.  This need for reversibility has many consequences for the
design of quantum gates.  Clearly for a gate to be reversible it must
be possible to reconstruct the input bits knowing only the design of the
gate and the output bits, and so every input bit must be in some sense
preserved in the outputs.  One trivial consequence of this is that the
gate must have exactly as many outputs as inputs.  For this reason it is
obvious that gates such as {\sc and} and {\sc or} are not reversible.
It is, however, possible to construct reversible equivalents of {\sc and}
and {\sc or}, in which the input bits are preserved.

Just as it can be shown that one or more gates (such as the {\sc nand} gate)
are universal for classical computing\cite{feynman2}
(that is, any classical gate can
be constructed using only wires and {\sc nand} gates), it can be shown
that certain gates or combinations of gates are universal for quantum
computing.  In particular it can be shown\cite{barenco}
that the combination of a
general single qubit rotation with the two bit ``controlled-{\sc not}''
gate ({\sc cnot}) is universal.  Furthermore it is possible to
build a reversible equivalent of the {\sc nand} gate, and thus to
implement any classical logic operation using reversible logic.

Single qubit rotations are easily implemented in NMR as they correspond
to rotations within the subspace corresponding to a single spin, and
such rotations can be achieved using radio frequency (RF) fields.  One
particularly important single bit gate is the Hadamard gate, which 
performs the rotational transformation
\begin{equation}
\begin{split}
\ket{0}&\stackrel{H}{\longrightarrow}\frac{\ket{0}+\ket{1}}{\sqrt{2}}\\
\ket{1}&\stackrel{H}{\longrightarrow}\frac{\ket{0}-\ket{1}}{\sqrt{2}}.
\end{split}
\end{equation}
The Hadamard operator can thus be used to convert eigenstates into
superpositions of states.  Similarly, as the Hadamard is self-inverse,
it can be used to convert superpositions of states back into eigenstates
for later analysis.

Two-bit gates correspond to rotations within subspaces
corresponding to two spins, and thus require some kind of spin--spin
interaction for their implementation.  In NMR the scalar spin--spin
coupling (J-coupling) has the correct form, and is ideally suited for
the construction of controlled gates, such as {\sc cnot}.  This gate
operates to invert the value of one qubit when another qubit (the
control qubit) has some specified value, usually \ket{1}; its truth
table is shown in table~\ref{table:cnot}.
\begin{table}
\begin{center}
\begin{tabular}{|l|l||l|l|}
\hline
\multicolumn{2}{|c||}{input} & \multicolumn{2}{|c|}{output} \\\hline
0 & 0 & 0 & 0 \\
0 & 1 & 0 & 1 \\
1 & 0 & 1 & 1 \\
1 & 1 & 1 & 0 \\\hline
\end{tabular}
\end{center}
\caption{The truth table for the {\sc cnot} gate.  The first qubit (the
control qubit) is unchanged by the gate, while the second qubit is flipped
if the control qubit is in state 1, effectively implementing an {\sc xor}
gate.}
\label{table:cnot}
\end{table}

Finally it is necessary to have some way of reading out
information about the final quantum state of the system, and thus
obtaining the result of the calculation.  In most implementations of quantum
computers this process is equivalent to determining which of two
eigenstates a two level system is in, but this is not a practical approach
in NMR.  It is, however, possible to obtain equivalent information by exciting
the spin system and observing the resulting NMR spectrum.  Different qubits
correspond to different spins, and thus give rise to signals at different
resonance frequencies, while the eigenstate of a spin before the excitation
can be determined from the relative phase (absorption or emission) of the
NMR signals.

\section{THE DEUTSCH ALGORITHM}
Deutsch's problem in its simplest form concerns the analysis of
single bit binary functions:
\begin{equation}
f(x):B\mapsto B,
\end{equation}
where $B=\{0,1\}$ is the set of possible values for a single bit.
Such functions take a single bit as input, and return a single bit as their
result.  Clearly there are exactly four such functions, which may be
described by their truth tables, as shown in table~\ref{table:f}.
\begin{table}
\begin{center}
\begin{tabular}{|l|l|l|l|l|}
\hline
$x$&$f_{00}(x)$&$f_{01}(x)$&$f_{10}(x)$&$f_{11}(x)$ \\\hline
0& 0 & 0 & 1 & 1\\
1& 0 & 1 & 0 & 1\\\hline
\end{tabular}
\end{center}
\caption{The four possible binary functions mapping one bit to another.}
\label{table:f}
\end{table}
These four functions can be divided into two groups: the two ``constant''
functions, for which $f(x)$ is independent of $x$ ($f_{00}$ and
$f_{11}$), and the two ``balanced'' functions, for which $f(x)$ is
zero for one value of $x$ and unity for the other ($f_{01}$ and $f_{10}$).
Given some unknown function $f$ (known to be one of these four functions),
it is possible to determine which of the four functions it is by applying
$f$ to two known inputs: $0$ and $1$.  This procedure also provides
enough information to determine whether the function is constant or
balanced.  However knowing whether the function is constant or balanced
corresponds to only one bit of information, and so it might be possible
to answer this question using only one evaluation of the function $f$.
Equivalently, it might be possible to determine the value of
$f(0)\oplus f(1)$ using only one evaluation of $f$.  (The symbol
$\oplus$ indicates addition modulo $2$, and for two one bit numbers, $a$
and $b$, $a\oplus b$ equals $0$ if $a$ and $b$ are the same, and $1$ if
they are different.)
In fact this can be achieved as long as the calculation is performed
using a quantum computer rather than a classical one.

Quantum computers of necessity use reversible logic, and so it is not
possible to implement the binary function $f$ directly.  It is, however,
possible to design a propagator, $U_f$, which captures $f$ within a reversible
transformation by using a system with two input qubits and two output qubits
as follows
\begin{equation}
\ket{x}\ket{y}\stackrel{U_f}{\longrightarrow}\ket{x}\ket{y\oplus f(x)}.
\end{equation}
The two input bits are preserved ($x$ is preserved directly, while $y$
is preserved by combining it with $f(x)$, the desired result), and so
$U_f$ corresponds to a reversible transformation.  Note that for
any one bit number $a$, $0\oplus a=a$, and so values of $f(x)$ can be
determined by setting the second input bit to $0$.  Using this propagator and
appropriate input states it is possible to evaluate $f(0)$ and $f(1)$
using
\begin{equation}
\ket{0}\ket{0}\stackrel{U_f}{\longrightarrow}\ket{0}\ket{f(0)}
\end{equation}
and
\begin{equation}
\ket{1}\ket{0}\stackrel{U_f}{\longrightarrow}\ket{1}\ket{f(1)}.
\end{equation}

\begin{figure}
\begin{center}
\begin{picture}(200,100)
\put(15,70){\makebox(0,0)[r]{\ket{x}}}
\put(20,70){\line(1,0){50}}
\put(15,30){\makebox(0,0)[r]{\ket{0}}}
\put(20,30){\line(1,0){50}}
\put(70,20){\framebox(20,60){$U_f$}}
\put(90,70){\line(1,0){50}}
\put(145,70){\makebox(0,0)[l]{\ket{x}}}
\put(90,30){\line(1,0){50}}
\put(145,30){\makebox(0,0)[l]{\ket{f(x)}}}
\end{picture}
\end{center}
\caption{Quantum circuit for the analysis of a binary function $f$.}
\label{fig:cf}
\end{figure}
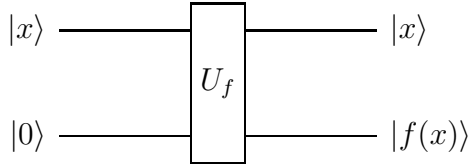

The approach outlined above, in which the state of a quantum computer is
described explicitly,
swiftly becomes unwieldy, and it is useful to use more compact notations.
One particularly simple approach is to use quantum circuits,\cite{deutsch4}
which may
be drawn by analogy with classical electronic circuits.  In this approach
lines are used to represent ``wires'' down which qubits ``flow'', while
boxes represent quantum gates which perform appropriate unitary
transformations.  For example, the analysis of
$f$ can be summarised by the circuit shown in figure~\ref{fig:cf}.

So far this is simply using a quantum computer to simulate a classical
computer implementing classical algorithms.  With a quantum computer,
however, it is not necessary to start with the system in some eigenstate;
instead it is possible to begin with a superposition of states.  Suppose
the calculation begins with the second qubit in the superposition
$(\ket{0}-\ket{1})/\sqrt{2}$.  Then
\begin{equation}
\begin{split}
\ket{x} \left(\frac{\ket{0}-\ket{1}}{\sqrt{2}}\right)
&\stackrel{U_f}{\longrightarrow}
\ket{x} \left(\frac{\ket{0\oplus f(x)}-\ket{1\oplus f(x)}}{\sqrt{2}}\right)\\
&=
\begin{cases} \displaystyle
\ket{x}\frac{\ket{0}-\ket{1}}{\sqrt{2}} &\text{if $f(x)=0$},\\
\displaystyle
\ket{x}\frac{\ket{1}-\ket{0}}{\sqrt{2}} &\text{if $f(x)=1$}
\end{cases}\\
&=(-1)^{f(x)}\:\ket{x}\frac{\ket{0}-\ket{1}}{\sqrt{2}}.
\end{split}
\end{equation}
(We have used the fact that $0\oplus a=a$, as before, while $1\oplus a=1$
if $a=0$ and $0$ if $a=1$).
The value of $f(x)$ is now encoded in the overall phase of the result,
with the qubits left otherwise unchanged.
While this is not particularly useful, suppose the calculation begins
with the first qubit also in a superposition of states, namely
$(\ket{0}+\ket{1})/\sqrt{2}$.  Then\cite{ekert2}
\begin{equation}
\begin{split}
\left(\frac{\ket{0}+\ket{1}}{\sqrt{2}}\right)
\left(\frac{\ket{0}-\ket{1}}{\sqrt{2}}\right)
&\stackrel{U_f}{\longrightarrow}
\left(\frac{(-1)^{f(0)}\:\ket{0}+(-1)^{f(1)}\:\ket{1}}{\sqrt{2}}\right)
\left(\frac{\ket{0}-\ket{1}}{\sqrt{2}}\right)\\
&=
(-1)^{f(0)}\left(\frac{\ket{0}+(-1)^{f(0)\oplus f(1)}\:\ket{1}}{\sqrt{2}}\right)
\left(\frac{\ket{0}-\ket{1}}{\sqrt{2}}\right)
\end{split}
\end{equation}
with the first qubit ending up in the superposition
$(\ket{0}\pm\ket{1})/\sqrt{2}$, with the desired answer ($f(0)\oplus f(1)$)
encoded as the \emph{relative} phase of the two states contributing to
the superposition.  This relative phase can be measured, and so the value
of $f(0)\oplus f(1)$ (that is, whether $f$ is constant or balanced) has been
determined using only one application of the propagator $U_f$, that is only
one evaluation of the function $f$.

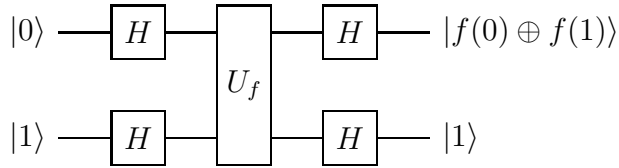
\begin{figure}
\begin{center}
\begin{picture}(210,100)
\put(15,70){\makebox(0,0)[r]{\ket{0}}}
\put(20,70){\line(1,0){20}}
\put(40,60){\framebox(20,20){$H$}}
\put(60,70){\line(1,0){20}}
\put(15,30){\makebox(0,0)[r]{\ket{1}}}
\put(20,30){\line(1,0){20}}
\put(40,20){\framebox(20,20){$H$}}
\put(60,30){\line(1,0){20}}
\put(80,20){\framebox(20,60){$U_f$}}
\put(100,70){\line(1,0){20}}
\put(120,60){\framebox(20,20){$H$}}
\put(140,70){\line(1,0){20}}
\put(165,70){\makebox(0,0)[l]{\ket{f(0)\oplus f(1)}}}
\put(100,30){\line(1,0){20}}
\put(120,20){\framebox(20,20){$H$}}
\put(140,30){\line(1,0){20}}
\put(165,30){\makebox(0,0)[l]{\ket{1}}}
\end{picture}
\end{center}
\caption{A quantum circuit for solving Deutsch's problem.}
\label{fig:cd}
\end{figure}

This approach can be easily implemented using a quantum circuit,
as shown in figure~\ref{fig:cd}.
This circuit starts off from appropriate eigenstates, uses Hadamard
transformations to convert these into superpositions, applies the
propagator $U_f$ to these superpositions, and finally use another pair
of Hadamard transforms to convert the superpositions back into eigenstates
which encode the desired result.

\section{IMPLEMENTING THE DEUTSCH ALGORITHM IN NMR}
The Deutsch algorithm can be implemented on a quantum computer with
two qubits, such as an NMR quantum computer based on two coupled
spins.  First it is necessary to show that the individual components of
the quantum circuit can be built.  It is convenient to begin by writing
down the necessary states and operators using the product operator
basis set\cite{ernst,sorensen} normally used in describing NMR experiments
(this basis is formed by taking outer products between Pauli matrices
describing the individual spins, together with the scaled unit matrix,
$\smhalf E$).

The initial state, $\ket{\psi_{01}}=\ket{0}\ket{1}$, can be written as
a vector in Hilbert space,
\begin{equation}
\ket{\psi_{01}}=\begin{pmatrix} 0 \\ 1 \\ 0 \\ 0 \end{pmatrix},
\end{equation}
but this description is not really appropriate.  Unlike other implementations,
an NMR quantum computer comprises not just a single set of spins but
rather an ensemble of spins in a statistical mixture of states.  Such
a system is most conveniently treated using a density matrix, which
can describe either a mixture or a pure state; for example
\begin{equation}
\rho_{01}=\ket{\psi_{01}}\bra{\psi_{01}}=\begin{pmatrix}
0 & 0 & 0 & 0 \\
0 & 1 & 0 & 0 \\
0 & 0 & 0 & 0 \\
0 & 0 & 0 & 0
\end{pmatrix}.
\end{equation}
This density matrix can be decomposed in the product operator basis as
$\rho_{01}=(I_z-S_z-2I_zS_z+\smhalf E)/2$.  Ignoring multiples of the unit
matrix (which give rise to no observable effects in any NMR experiment)
this can be reached from the
thermal equilibrium density matrix ($I_z+S_z$) by a series of RF
and field gradient pulses\cite{cory2}.

The unitary transformation matrix corresponding to the Hadamard operator
on a single spin can be written as
\begin{equation}
H=\frac{1}{\sqrt{2}}\begin{pmatrix} 1 & 1 \\ 1 & -1 \end{pmatrix}.
\end{equation}
This corresponds to a $180^\circ$ rotation around an axis tilted at
$45^\circ$ between the $z$ and $x$-axes.  Such a rotation can be achieved
directly using an off resonance pulse,\cite{ernst} or using a three
pulse sandwich\cite{ernst} such as $45^\circ_y - 180^\circ_x - 45^\circ_{-y}$.
Even more simply the Hadamard can be approximated by a $90^\circ_y$ pulse.
While this is clearly not a true Hadamard operator (for example, it is
not self inverse), its behaviour is similar and it can be used in some
cases: for example, it is possible to replace the first pair of Hadamard
gates in the circuit for the Deutsch Algorithm (figure~\ref{fig:cd}) by
$90^\circ_y$ pulses and the second pair of gates by $90^\circ_{-y}$ pulses.
Clearly it is possible to apply the Hadamard operator either to just one of
the two spins (using selective soft RF pulses\cite{freeman1}) or to both spins
simultaneously (using non-selective hard pulses).

The unitary transformations corresponding to the four possible propagators
$U_f$ are also easily derived.  Each propagator corresponds to flipping
the state of the second qubit under certain conditions as follows:
$U_{00}$, never flip the second qubit; $U_{01}$, flip the second qubit
when the first qubit is in state one; $U_{10}$, flip the second qubit
when the first qubit is in state zero; $U_{11}$, always flip the second
qubit.  The first and last cases are particularly simple, as $U_{00}$
corresponds to doing nothing (the identity operation), while $U_{11}$
corresponds to inverting the second spin (a conventional {\sc not} gate,
or, equivalently, a $180^\circ$ pulse).  The
second and third propagators correspond to controlled-{\sc not} gates,
which can be implemented using spin--spin couplings.  For example
$U_{01}$ is described by the matrix
\begin{equation}
U_{01}=\begin{pmatrix}
1 & 0 & 0 & 0 \\
0 & 1 & 0 & 0 \\
0 & 0 & 0 & 1 \\
0 & 0 & 1 & 0
\end{pmatrix},
\end{equation}
which can be achieved using the pulse sequence
\begin{equation}
90\,S_y - \text{\emph{couple}} - 90\,I_z - 90\,S_{-z} - 90\,S_{-y}
\end{equation}
where $90\,S_y$ indicates a $90^\circ$ pulse on the second spin, 
\emph{couple} indicates evolution under the scalar coupling Hamiltonian,
$\pi J_{IS}\,2I_zS_z$, for a time $1/2J_{IS}$, and $90\,I_z$ and $90\,S_{-z}$
indicate either periods of free precession under Zeeman Hamiltonians
or the application of composite $z$-pulses.\cite{freeman1,freeman2}
Similarly $U_{10}$ can be achieved using the pulse sequence
\begin{equation}
90\,S_y - \text{\emph{couple}} - 90\,I_z - 90\,S_z - 90\,S_{-y}.
\end{equation}

The pulse sequences described above can be implemented in many different
ways, as different composite $z$-pulses can be used, the order of
some of the pulses can be varied, and in some cases different pulses can
be combined together.  We chose to use the implementation
\begin{equation}
90\,S_y - 1/4J_{IS} - 180_x - 1/4J_{IS} - 180_x - 90\,I_y - 90\,I_x -
90_{-y} - 90\,S_{\pm x}
\end{equation}
where pulses not marked as either $I$ or $S$ were applied to both nuclei.  
The phase of the final pulse distinguishes $U_{01}$ (for which the final
pulse is $S_{+x}$) from $U_{10}$ (for which it is $S_{-x}$).

\begin{figure}
\begin{center}
\begin{picture}(230,100)
\put(0,90){\makebox(0,0){(a)}}
\put(15,70){\makebox(0,0)[r]{\ket{0}}}
\put(20,70){\line(1,0){80}}
\put(15,30){\makebox(0,0)[r]{\ket{0}}}
\put(20,30){\line(1,0){80}}
\put(100,20){\framebox(20,60){$U_f$}}
\put(120,70){\line(1,0){25}}
\put(145,60){\framebox(30,20){$90^\circ_{y}$}}
\put(175,70){\line(1,0){25}}
\put(205,70){\makebox(0,0)[l]{$+x$}}
\put(120,30){\line(1,0){25}}
\put(145,20){\framebox(30,20){$90^\circ_{y}$}}
\put(175,30){\line(1,0){25}}
\put(205,30){\makebox(0,0)[l]{$\pm x$}}
\end{picture}
\end{center}
\begin{center}
\begin{picture}(230,100)
\put(0,90){\makebox(0,0){(b)}}
\put(15,70){\makebox(0,0)[r]{\ket{0}}}
\put(20,70){\line(1,0){25}}
\put(45,60){\framebox(30,20){$90^\circ_y$}}
\put(75,70){\line(1,0){25}}
\put(15,30){\makebox(0,0)[r]{\ket{1}}}
\put(20,30){\line(1,0){25}}
\put(45,20){\framebox(30,20){$90^\circ_y$}}
\put(75,30){\line(1,0){25}}
\put(100,20){\framebox(20,60){$U_f$}}
\put(120,70){\line(1,0){80}}
\put(205,70){\makebox(0,0)[l]{$\pm x$}}
\put(120,30){\line(1,0){80}}
\put(205,30){\makebox(0,0)[l]{$-x$}}
\end{picture}
\end{center}
\caption{Modified quantum circuits for the analysis of binary functions
on an NMR quantum computer.  (a) A circuit for the classical analysis
of $f(0)$; the normal circuit (see figure~\ref{fig:cf}) is followed by
$90^\circ_y$ pulses to excite the NMR spectrum.  Clearly $f(1)$ can be
obtained in a similar way.  (b)  A circuit for the implementation of
the Deutsch algorithm, with Hadamard operators replaced by
$90^\circ_{\pm y}$ pulses.  The final $90^\circ_y$ excitation pulses
cancel out the $90^\circ_{-y}$ pulses, and thus all four pulses can be
omitted.}
\label{fig:cdnmr}
\end{figure}
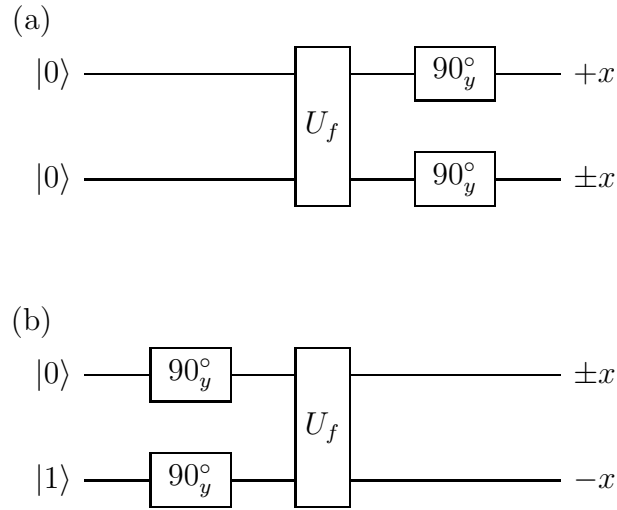

Finally it is necessary to consider analysis of the final state, which
could in general be one of the four states $\rho_{00}$, $\rho_{01}$,
$\rho_{10}$, or $\rho_{11}$.  In order to distinguish these states
it is necessary to apply a $90^\circ_y$ pulse and observe the NMR spectrum.
The final NMR signal observed from spin $I$ is $I_x$ if the spin
is in state $0$, and
$-I_x$ if it is in state $1$.  For a computer implementing the Deutsch
algorithm the final detection $90^\circ_y$ pulses cancel out the two
final pseudo-Hadamard $90^\circ_{-y}$ pulses, and thus all four pulses
can be omitted (see figure~\ref{fig:cdnmr}).
The final NMR signal observed is either $\smhalf I_x-\smhalf S_x$
(corresponding to $\rho_{01}$) or $-\smhalf I_x-\smhalf S_x$
(corresponding to $\rho_{11}$).  Hence it
is simple to determine the value of $f(0)\oplus f(1)$ (that is, determine
whether the function is constant or balanced) by determining the relative
phase of the signals from the two spins.

\section{EXPERIMENTAL RESULTS}
In order to demonstrate the results described above, we have constructed
an NMR quantum computer capable of implementing the Deutsch algorithm.  For
our two-spin system we chose to use a $\rm50\,mM$ solution of
the pyrimidine base cytosine in
$\rm D_2O$; rapid exchange of the two amine protons and
the single amide proton with the deuterated solvent leaves two
remaining protons forming an isolated two-spin system.
All NMR experiments were conducted at $\rm20^\circ C$ and $\rm pH^*=7$
on a home-built
NMR spectrometer at the Oxford Centre for Molecular Sciences, with a
\nuc{1}{H} operating frequency of \MHz{500}.  The observed J-coupling
between the two protons was \Hz{7.2}, while the difference in resonance
frequencies was \Hz{763}.  Selective excitation was achieved using
Gaussian\cite{bauer} soft pulses incorporating a phase ramp\cite{geen,kupce}
to allow excitation away from the transmitter frequency.
During a selective pulse the other (unexcited) spin continues
to experience the main Zeeman interaction, resulting in a rotation around
the $z$-axis, but the length of the selective pulses can be chosen such that
the net rotation experienced by the other spin is zero.  The residual
HOD resonance was suppressed by low power saturation during the relaxation
delays.

\begin{figure}
\begin{center}
\epsfig{file=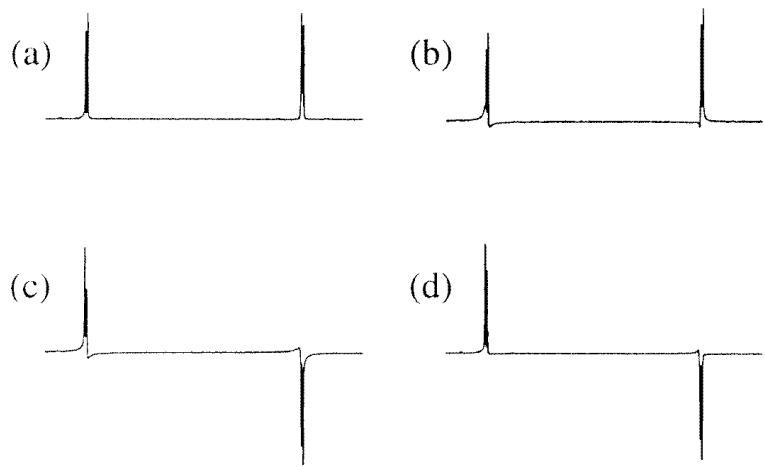,width=100mm}
\end{center}
\caption{Experimental implementation of an algorithm to determine $f(0)$
on an NMR quantum computer.  (a)  The result of applying $U_{f_{00}}$; as
this propagator is the identity matrix this spectrum can also serve as a
reference.  The lefthand pair of signals corresponds to the first spin ($I$),
while the pair on the right hand side correspond to the second spin ($S$).
Note that the signals from both spins (which are in state \ket{0}, the ground
state) are in absorption.  (b) The result of applying $U_{f_{01}}$; both sets
of signals are still in absorption, as $f(0)=0$ for this function.
(c)  The result of applying $U_{f_{10}}$; the signals from spin $S$ are
now in emission, since $f(0)=1$ for this function.  (d) The result of
applying $U_{f_{11}}$; the signals from spin $S$ are once again in emission
as expected.}
\label{fig:rf0}
\end{figure}

This system can be used both for the implementation of classical algorithms
to analyse $f(0)$ and $f(1)$ and for the implementation of the Deutsch
algorithm; as shown in figure~\ref{fig:cdnmr} the pulse sequences differ
only in the placement of the $90^\circ_y$ pulses.
The results for the classical algorithm to determine $f(0)$ are shown in
figure~\ref{fig:rf0}.
The lefthand pair of signals corresponds to the first spin ($I$),
while the pair on the right hand side correspond to the second spin ($S$);
the (barely visible) splitting in each pair arises from the scalar coupling
$J_{IS}$.  In this experiment the value of $f(0)$ is determined by setting
both spins $I$ and $S$ into state \ket{0}, performing the calculation, and
then measuring the final state of spin $S$; spin $I$ should not be
affected, and so should remain in state \ket{0}.  The phase of the
reference spectrum (a) was adjusted so that signals from spin $I$ appear in
absorption, and the same phase correction was applied to the other three
spectra.  The state of a spin after a calculation can then be determined
by determining whether the corresponding signals in the spectrum are
in absorption (state \ket{0}) or emission (state \ket{1}).  As expected
spin $I$ does indeed remain in state \ket{0}, while the value of $f(0)$
(determined from spin $S$) is $0$ for $U_{f_{00}}$ and $U_{f_{01}}$
but $1$ for $U_{f_{10}}$ and $U_{f_{11}}$.

\begin{figure}
\begin{center}
\epsfig{file=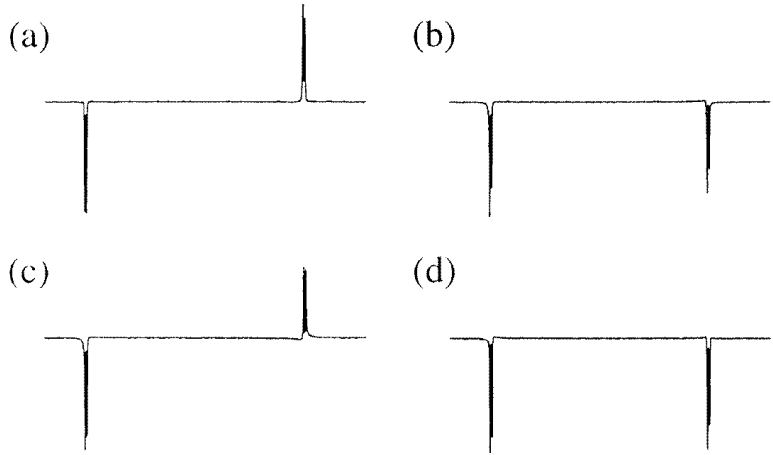,width=100mm}
\end{center}
\caption{Experimental implementation of an algorithm to determine $f(1)$
on an NMR quantum computer; in this case the algorithm starts with spin $I$
in the excited state, \ket{1}, and so signals from spin $I$ are in emission.
For details of the labelling see figure~\ref{fig:rf0}.}
\label{fig:rf1}
\end{figure}

Clearly our NMR quantum computer is capable of implementing this classical
algorithm, as it is simple to determine $f(0)$.  The other value, $f(1)$,
can be determined in a very similar way (see figure~\ref{fig:rf1}).
In this case spin $I$ remains in state \ket{1}, while $f(1)$ equals $0$ for
$U_{f_{00}}$ and $U_{f_{10}}$ and equals $1$ for $U_{f_{01}}$ and
$U_{f_{11}}$.
There are, however, several imperfections visible in the results.

First, the signals are not perfectly phased: rather than exhibiting
pure absorption or pure emission lineshapes the signals have more
complex shapes, including dispersive components.  These arise
from the difficulty of implementing perfect selective pulses, which
effect the desired rotation at one spin while leaving the other spin
entirely unaffected.  Similarly the selective pulses will not perfectly
suppress J-couplings during the excitation, leading to the appearance of
antiphase contributions to the lineshape.  Any practical selective pulse
will be imperfect, and so will result in systematic distortions in the
final result.  Note that these distortions are most severe in cases
(b) and (c), where the propagator is complex, containing a large number of
selective pulses.  Interestingly the distortions are also more severe
for the measurement of $f(0)$ (figure~\ref{fig:rf0}) than for $f(1)$
(figure~\ref{fig:rf1}); there is no simple explanation for this effect,
which is due to the complex interplay of many selective pulses.  We are
currently seeking ways to minimise these effects.

Secondly, the signal intensities vary in different cases; as before the
signal loss is most severe in cases (b) and (c), corresponding to complex
propagators.  This is in part a consequence of imperfect selective pulses,
as discussed above, but may also indicate the effects of spin
relaxation, that is decoherence of the states involved in the calculation.
Decoherence is a fundamental problem, and may ultimately limit the size of
practical quantum computers,\cite{chuang3,plenio1,plenio2} although a variety
of error correction techniques\cite{shor2,steane1,steane2} have been devised
to overcome it.

These imperfections are not a major problem in our NMR quantum computer,
as it is still easy to determine the state of a spin.  However our
computer is small, and the programs run on it are short (that is,
they contain a small number of logic gates); if more complex programs are
to be run on larger computers then these imperfections must be addressed.

\begin{figure}
\begin{center}
\epsfig{file=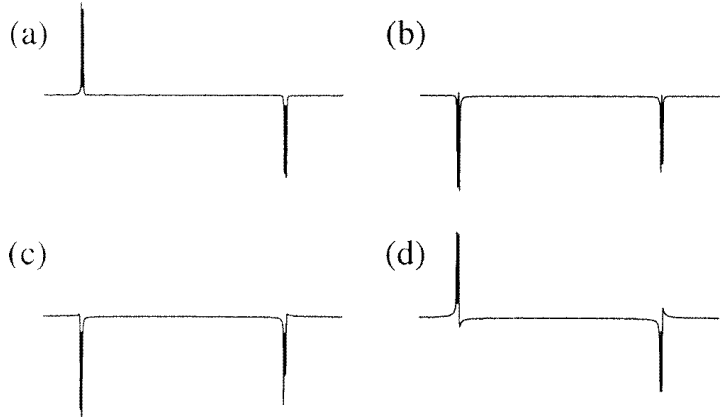,width=100mm}
\end{center}
\caption{Experimental implementation of a quantum algorithm to
determine $f(0)\oplus f(1)$ on an NMR quantum computer.  In this case the
result can be read out on spin $I$, that is using the signals on
the left of the spectrum.  For details of the labelling see
figure~\ref{fig:rf0}.  As expected the $I$ spin is inverted when the
function is balanced ($f_{01}$ or $f_{10}$), but not when the function
is constant ($f_{00}$ or $f_{11}$).}
\label{fig:rd}
\end{figure}

The results of implementing the Deutsch quantum algorithm are shown in
figure~\ref{fig:rd}.
In this case the result ($f(0)\oplus f(1)$) can be read from the final state
of spin $I$, while spin $S$ remains in state \ket{1}.  As expected spin $I$
is in state \ket{0} for the two constant functions ($f_{00}$ and $f_{11}$),
but in state \ket{1} for the two balanced functions ($f_{01}$ and $f_{10}$).
Once again a number of imperfections are visible, though in this case they
appear to be most severe in the case of $U_{f_{11}}$.

\section{SUMMARY}
We have demonstrated that the isolated pair of \nuc{1}{H} nuclei in
partially deuterated cytosine can be used to implement a two qubit NMR
quantum computer.  This computer can be used to run both classical
algorithms and quantum algorithms, such as that for solving Deutsch's
problem (distinguishing between constant and balanced functions).  This
is the first successful implementation of a quantum algorithm on any
physical system\cite{update}.

This result confirms that NMR shows great promise as a technology for
the implementation of small quantum computers.  Difficulties do exist,
largely as a result of the large number of selective pulses involved
in the implementation of quantum gates, but we are currently seeking
ways to overcome these problems.  Even with the current level of errors
it should be possible to build a three qubit computer capable of
implementing more complex logic gates and algorithms.

\section*{ACKNOWLEDGEMENTS}
We are indebted to R.~H. Hansen (Clarendon Laboratory) for invaluable
advice and assistance.
We thank N.~Soffe and J.~Boyd (OCMS) for assistance with implementing the
NMR pulse sequences.  We are grateful to A.~Ekert (Clarendon Laboratory) and
R.~Jozsa (University of Plymouth) for helpful conversations.  JAJ thanks
C.~M. Dobson (OCMS) for his encouragement and support.  This is a contribution
from the Oxford Centre for Molecular Sciences which is supported by the UK
EPSRC, BBSRC and MRC.  MM thanks CESG (UK) for their support.

\newpage


\begin{thebibliography}{10}
\bibitem{update}
Since initial submission of this manuscript there has
been considerable progress in this field, including another implementation
of an algorithm to solve Deutsch's problem\cite{chuang4}, and two
implementations of Grover's quantum search
algorithm\cite{chuang5,jones1,jones2}.
\end{thebibliography}
\end{document}